\documentstyle[aps,preprint,eqsecnum]{revtex}
\begin{document}
\thispagestyle{empty}
\title{Symplectic structure of electromagnetic duality}
\author{M. Kachkachi}
%\date{}

\address{
D\'epartement de Math\'ematiques, F.S.T.S.,B.P. 577,\\
Universit\'e Hassan 1$^{\rm er}$, Settat, Morocco\thanks{Permanent Address}\\
and\\
UFR-HEP-Rabat, Universit\'e MV, Facult\'e des Sciences, \\
D\'epartement de Physiques,
B.P. 1400, Rabat, Morocco}
\maketitle

\begin{abstract}
We develop an elecromagnetic symplectic structure on the space-time
manifold by defining a Poisson bracket in terms of an invertible
electromanetic tensor $F_{\mu \nu}$. Moreover, we define electromagnetic
symplectic diffeomorphisms by imposing a special condition on the
electromagnetic tensor. This condition implies a special constraint on
the Levi-Civita tensor. Then, we express geometrically the
electromagnetic duality by defining a dual local coordinates system.
\end{abstract}
\newpage
\section{ Introduction}
In the last twenty years, a field-theoretical Poisson geometry has been
developed and was proved to be useful for the study of integrable
systems described in terms of ordinary differential equations, partial
differential equations and quantum field theory [1].

Symplectic groups have been shown to be powerful material to handle
symmetries of quantum field theories: the group $SDiff(S^{2})$ of the
symplectic transformations on the sphere $S^{2}$ is a residual symmetry
in relativistic membrane when quantized in the light cone gauge [2]. The
reformulation of $D=4 \ SU(\infty)$ Yang-Mills theory as a Poisson gauge
theory on the space $M_{6} = M_{4}\bigotimes S^{2}$, where $M_{4}$ is
the Minkowski space -time, was based  on the equivalence
$SDiff(S^{2})\sim SU(N\rightarrow \infty)$ [3]. Moreover, the developement of
the symplectic structure of harmonic superspace ( which is the framework
of an off-shell formulation of $D = 4$ $N =  2$ supersymmetric theories)
has been proved  useful to derive a general form of the symplectic
invariant coupling of the Maxwell gauge prepotential [4].

An other interesting application of symplectic geometry in physics was
given by Guillemin and Sternberg [5], where they have proved that the
electromagnetic field ( and charges ) determines the symplectic
structure of the eight-dimensional phase space of a relativistic
particle. this relationship between electromagnetism and space-time is,
on one hand a pure relativistic phenomenon and on the other hand a
certain kind of duality. Indeed, duality has appeared in different
contexts in theoretical physics and can be used as an organizing
principle [6]. It means that there is two equivalent descriptions of a
model using different fields which are related by  Legendre
transformations [7]. One early example is the $4D$ duality between
electricity and magnetism in Maxwell,s theory that exchanges the weak
coupling regime with the strong coupling one. A more sophisticated kind of
duality was conjectured by Montonen and Olive [8] to hold for
non-abelian gauge theories which interchanges electric\\ charges ( related
via Noether,s theorem to the existence of symmetries ) and magnetic
charges which are topological in nature. Seiberg and Witten have
produced a supersymmetric version of these ideas [9]. Very recently, Sen
has conjectured that such duality implies results about
$L^{2}$-cohomology of monopole moduli spaces.

In this paper we establish a symplectic structure on the space-time
manifold via the electromagnetic tensor $F_{\mu \nu}$ of the Maxwell theory.
Indeed, an "electromagnetic Poisson structure" on the symplectic
electromagnetic manifold $ ( F,M_{4},\Gamma)$, where $ F$ is the
elecromagnetic 2-form, $M_{4}$ is the space-time manifold and $\Gamma$
is the stationary surface: the space of all gauge fields that satisfy
equations of motion and for which the matrix $ (F_{\mu \nu})$ is
invertible. Moreover, we derive symplectic diffeomorphisms via a
specific constraint on the electromagnetic tensor. Then, we express
geometrically the electromagnetic duality by giving a "dual local
coodinates system" of the first one where the symplectic electromagnetic
structure is defined.

In section 2 we recall the symplectic manifold properties that is endowed
with a Poisson bracket structure. In section 3 we construct a symplectic
structure where the symplectic form is the electromagnetic tensor. Then, we define
electromagnetic symplectic diffeomorphisms by imposing a specific
constraint on the electromagnetic tensor which induces a new relation
for the Levi-Civita tensor. In section 4 we give a symplectic form to
the electromagnetic field equations of motion.. In section 5 we
establish a geometrical structure of the electromagnetic duality by
expressing the constraint duality in terms of the diffeomorphisms
operators. Section 6 is devoted to our conclusion and open problems.
\section { Symplectic structure of a symplectic manifold}
An even dimensional manifold $M^{2n}$ endowed with a closed 2-form
$\omega$, i.e. $d\omega= 0$, is called a symplectic manifold and the
structure given by $\omega$ is the symplectic structure. The
antisymmetric and non-degenerate bilinear 2-form $\omega$ is locally
written as [1] :
\begin{eqnarray}
\omega &=& {1\over2} \omega_{ij} dx^{i}\wedge dx^{j},  \ i,j =
1,...,2n\nonumber\\
det(\omega_{ij}) &\neq& 0,
\end{eqnarray}
where $(x^{i}),  i = 1, ..., 2n$ is a local coordinates system of $M^{2n}$.
This symplectic form defines a skew-symmetric scalar product on the
tangent space: for any two vectors $V = (V^{i})$ and $W = (W^{i})$ this
scalar product is given by
\begin{eqnarray}
(V,W) &=& \omega_{ij} V^{i}W^{j}\nonumber\\
&=& -(W,V).
\end{eqnarray}
The inverse matrix $(\omega^{ij})$ of $(\omega_{ij})$ is defined by
\begin{equation}
\omega^{ij}\omega_{ij} = \delta^{i}_{j}.
\end{equation}
If the symplectic form $\omega$ is exact, i.e. $\omega = d\theta$, where
$\theta = \theta_{k}(x) dx^{k}$ is a 1-form we get
\begin{equation}
\omega_{kl} = \partial_{k}\theta_{l} - \partial_{l}\theta_{k}.
\end{equation}
The symplectic manifold $ (M^{2n},\omega)$ can be interpreted as the
phase space of a classical system, where the obsevables have a natural
Poisson-Lie structure [12]. The Hamiltonian vector field $ V_{f} =
(V^{i}_{f})$ associated to an observable $f(x)$ is defined by the
equation
\begin{eqnarray}
i_{v_{f}}\omega & = & -  df\nonumber\\
               & \equiv & V^{i}_{f}\omega_{ij} dx^{i}
\end{eqnarray}
from which we get the vector field components:
\begin{equation}
V^{i}_{f} = \omega^{ij} \partial_{j}f.
\end{equation}
This Hamiltonian vector field generates a one-parameter family of local
canonical transformations. The Poisson bracket of two observables $f$
and $g$ is given in terms of their gradients
\begin{equation}
\{f,g\} = \omega^{ij}\partial_{i}f\partial_{j}g
\end{equation}
and satisfies the following requirements
\begin{eqnarray}
\{f,g\} &=& - \{g,f\}\nonumber\\
\{f+g,h\} &=& \{f,h\}\nonumber\\
\{ f g, h \} &=& f\{g,h\} + \{f,h\}g.
\end{eqnarray}
By using the coordinates $(x^{i})$, eq.(2.7) reduces to
\begin{equation}
\{x^{i},x^{j}\} =  \omega^{ij}.
\end{equation}
Furthermore, the Jacobi identity is equivalent to [1]:
\begin{equation}
\partial_{k}\omega_{ij}+ \partial_{j}\omega_{ki}+
\partial_{i}\omega_{jk} = 0, \forall i, j, k = 1, ..., 2n
\end{equation}
which is deduced from the equation
\begin{equation}
d({1\over2}\omega_{ij}dx^{i}\wedge dx^{j}) = 0.
\end{equation}
By definition, a function $f\in C^{\infty}(M^{2n})$ is a Casimir for the
Poisson bracket (2.7) if it belongs to its kernel, i.e. if for any
function $g\in C^{\infty}(M^{2n})$ we have [12]:
\begin{equation}
\{f,g\} = 0.
\end{equation}
The symplectic diffeomorphisms on the manifold $M^{2n}$ which are
connected to the identity and denoted by $Diff_{0}(M^{2n})$ are generated
by the operators [13]
\begin{equation}
L_{f} = \omega^{ij}\partial_{j}f\partial_{i},
\end{equation}
where $f$ is any arbitrary function  on $M^{2n}$. They obey the
following algebra
\begin{equation}
\ [  L_{f},L_{g} \ ] = L_{\{f, g \}}.
\end{equation}
\section{ Electromagnetic symplectic manifold}
Here, we construct a symplectic structure on the space-time manifold
whose symplectic form is the electromagnetic tensor $F_{\mu \nu}$ which
verifies all the symplectic properties of section 2. In this framework
we give a symplectic expression for the Maxwell,s equations and for the
electromagnetic Lagrangian [14].

 \subsection{ Electromagnetic Poisson structure}
Let us consider the electromagnetic field strength in the four Minkowski
space-time:
\begin{equation}
F = {1\over2} F_{\mu \nu}dx^\mu\wedge dx^\nu,
\end{equation}
 where
$$
F_{\mu \nu} = \partial_\mu A_\nu - \partial_{\nu} A_{\mu}
$$
is the antisymmetric tensor and $A_{\mu}$ is the electromagnetic field.
Furthermore, it is well known [15,16] that the 2-form $F$ is closed and
exact:
\begin{eqnarray}
dF = 0, \nonumber\\
F = dA,
\end{eqnarray}
where $A$ is the 1-form gauge field; $A(x) =  A_{\mu}(x) dx^{\mu}$.

Now, we define the electromagnetic symplectic structure as follows:
For any functions $f, g \in C^{ \infty} (M_{4})$ the Poisson bracket is given by
\begin{equation}
\{f,g\} = F^{\mu \nu} \partial_{\mu} f\partial_\nu g.
\end{equation}
In particular, in the local coordinates system $(x^{\mu}), \mu =
0,...,3$, we have
\begin{equation}
\{x^{\mu},x^{\nu}\} = F^{\mu \nu}.
\end{equation}
This equation (3.4) can be seen as a relationship between space-time and
electromagnetism in order to be endowed with a symplectic structure. The
corresponding properties are discussed in the next subsection.

The gauge fields considered here are elements of gauge orbits denoted by
$\Gamma$ [11], where $detF_{\mu \nu} \neq 0$ and  does not contain pure
gauge fields, i.e. $A_{\mu} = g^{-1} \partial_{\mu} g \Leftrightarrow
F_{\mu \nu} = 0$ [15]. On the other hand, the determinant of the
antisymmetric tensor $F_{\mu \nu}$ is given by
\begin{eqnarray}
det(F_{\mu \nu}) &=& ( {1 \over 8} \varepsilon^{\mu \nu \sigma \lambda} F_{\mu \nu} F_{ \sigma
\lambda})^2\nonumber\\
& = & ( {1\over 4} F_{\mu \nu}  \ ^*F^{\mu \nu})^2,
\end{eqnarray}
where $ ^*F^{\mu \nu} = {1\over 2} \varepsilon^{\mu \nu \sigma \lambda}
F_{\sigma \lambda}$ is the dual tensor of $F^{\mu \nu}$ in four dimensional
space-time. So, we consider the electromagnetic theory with a non
vanishing topological charge
\begin{eqnarray}
Q &=& {1\over 8\pi^{2}} \int F\wedge F\nonumber\\
  &=& {-1\over 4} \int C_{2}(A),
\end{eqnarray}
where $C_{2}(A)$ is the second Chern class [15].
Furthermore, we define the inverse of the matrix $(F^{\mu \nu})$ by
the relation
\begin{eqnarray}
F_{\mu \nu}\ ^*F^{\nu \sigma} &=& {1\over 4} \delta^{\sigma}_{\mu}(F_{\mu
\nu}\  ^*F^{\nu \mu})\nonumber\\
&=& \delta^{\sigma}_{\mu} \sqrt {det(F_{\mu \nu})}
\end{eqnarray}
which can be rewritten as
\begin{equation}
F_{\mu \nu} \ { ^*F^{\nu \sigma}\over \sqrt {det(F_{\mu \nu})}} =
\delta^{\sigma}_{\mu}.
\end{equation}
The Jacobi identity of this symplectic structure is the Bianchi identity
verified by the electromagnetic tensor:
\begin{equation}
\partial_\sigma F_{\mu\nu}+\partial_\nu F_{\sigma\mu}+
\partial_\mu F_{\nu\sigma}=0
\end{equation}
which is induced from the closure of the electromagnetic 2-form, i.e.
$ dF = 0 $.

\subsection{ Properties of the electromagnetic symplectic structure}

The electromagnetic field is a function of the space -time coordinates:
$ A_{\mu} = A_{\mu}(x) $. \
Then, Let us consider the particular solution for $ A_{\mu}$ that is,
\begin{equation}
A^{\mu}(x) = x^{\mu}.
\end{equation}
In this case the electromagnetic tensor $F_{\mu \nu}$ vanishes and the
eq.(3.4) leads to the constraint
\begin{equation}
\{x^{\mu},x^{\nu}\} = 0
\end{equation}
which means that these particular solutions (3.10) are observable
Casimirs of the electromagnetic symplectic structure. It means also that the
absence of electromagnetism is expressed by the triviality of the Poisson
bracket.

Now, using eq.(3.3) for the electromagnetic field $A^{\mu}$, i.e.
\begin{equation}
\{A^{\mu},A^{\nu}\} = F^{\sigma \lambda}\partial_{\sigma}
A^{\mu}\partial_{\lambda}A^{\nu}
\end{equation}
we get
\begin{equation}
\{A^{\mu},A^{\nu}\} =
L^{\mu}(A^{\nu}) - L^{\nu}(A^{\mu}) + A(A^{\mu})\partial_{\lambda}
\partial^\lambda A^{\nu} - A(A^{\nu})\partial_{\lambda}
\partial^{\lambda} A^{\mu},
\end{equation}
where
\begin{eqnarray*}
L^{\mu}&\equiv& \partial^{\lambda}(A^{\sigma} \partial_{\lambda}
A^{\mu})\partial_{\sigma},\nonumber\\
A(A^{\mu})&\equiv& A^{\sigma}\partial_\sigma A^{\mu}.
\end{eqnarray*}
Then, by considering equations of motion of the gauge field in the
Lorentz gauge $(\partial_{\mu} A^{\mu} = 0)$, the relation (3.13) reduces
to
\begin{eqnarray}
\{A^\mu,A^\nu\}&=&L^\mu(A^\nu)-L^\nu(A^\mu)\nonumber\\
&\equiv&[L,A]^{\mu \leftrightarrow \nu} \nonumber\\
&\equiv&-L^{\nu\mu}
\end{eqnarray}
The eq.(3.14) tells us that the electromagnetic symplectic structure is
realized on gauge orbits as a commutation relation. On the other hand, if
we consider the particular solution for $ A^{\mu} $ eq.(3.10), the
generators $L^{\mu}$ reduces to
\begin{equation}
L^{\mu} = \partial ^{\mu},
\end{equation}
which are tangent vectors on the space-time manifol, and  eq.(3.14)
coincides with  eq.(3.11). For the general case we get the following
commutation relations:
\begin{equation}
[L^{\mu},L^{\nu}] = \{\partial_{\sigma} (\partial^{\lambda} A^{\sigma}
\partial^{\rho} A^{m} \partial_{\rho} A^{\nu}) \partial_{\lambda}
A^{\mu}- \partial_{\sigma}( \partial^{\lambda} A^{\sigma} \partial^{\rho}
A^{m} \partial_{\rho} A^{\mu}) \partial_{\lambda} A^{\nu} \}
\partial_{m}
\end{equation}
\subsection{ Electromagnetic symplectic diffeomorphisms}

In analogy with the definition of symplectic diffeomorphisms given in
section 2 we define the generators of electromagnetic symplectic
diffeomophisms as follows:
\begin{equation}
L_{f} = F^{\mu \nu} \partial_{\nu} f \partial_{\mu},
\end{equation}
where $f$ is an arbitrary fuction of $x$. In particular, the associated
diffeomorphism generators to local coodinates $(x^{\lambda})$ are given
by
\begin{equation}
L_{x^{\lambda}} = F^{\mu \lambda} \partial_{\mu}.
\end{equation}
Furthermore, we recover the relation (3.4) as follows:
\begin{eqnarray}
L_{x^\mu}(x^\nu)&=&F^{\nu\mu}\nonumber\\
&=&\{x^\nu,x^\mu\}
\end{eqnarray}
So, we can verify that
\begin{equation}
L_{x^{\lambda}}(F^{\mu \nu}) = \{F^{\mu \nu},x^\lambda\},
\end{equation}
and
\begin{equation}
L_{x^{\lambda}}(A^\mu) = \{A^\mu,x^\lambda\}.
\end{equation}
So, the genertors $L_{x^{\lambda}}$ are realized as a Poisson bracket on
the observables:
\begin{eqnarray}
L_{x^{\lambda}} &=& \{x^\mu,x^\lambda\} \partial_{\mu}\nonumber\\
&=& \{,x^\lambda \}.
\end{eqnarray}
In order to have an electromagnetic symplectic diffeomorphism algebra we
impose the following commutation relations
\begin{eqnarray}
[L_{x^{\mu}},L_{x^{\nu}}] &=& L_{ \{ x^\mu,x^\nu\}}\nonumber\\
&=& L_{F^{\mu \nu}}
\end{eqnarray}
which can be rewritten, by considering the eq.(3.22), as
\begin{equation}
[\{,x^\mu \},\{,x^\nu\}] = \{,\{x^\mu,x^\nu\}\}.
\end{equation}
This means that the electromagnetic symplectic diffeomorphisms algebra
is realized as the Poisson bracket structure.

After some developments we find
\begin{equation}
L_{F^{\mu \nu}} = \partial_\lambda (F^{\sigma
\lambda} F^{\mu \nu})\partial_{\sigma}
\end{equation}
\begin{equation}
[L_{x^\mu},L_{x^\nu}] = \partial_\lambda (F^{\lambda \mu} F^{\sigma
\nu} - F^{\lambda \nu} F^{\sigma \mu}) \partial_\sigma.
\end{equation}
Then, from eq.(3.23) we get the following constraint on the
electromagnetic tensor up to a divergence term:
\begin{equation}
F^{\lambda \mu} F^{\sigma \nu} - F^{\lambda \nu} F^{\sigma \mu} =
F^{\sigma \lambda} F^{\mu \nu}.
\end{equation}
It is easy to verify that  the eq.(3.27) is invariant under a circular
permutation of indices $(\lambda ,\mu ,\sigma ,\nu )$. On the other
hand, by expressing eq.(3.27) in terms of the dual tensor \ $^*F^{\mu
\nu} = {1\over2}\varepsilon^{\mu \nu \sigma \lambda} F_{\sigma
\lambda}$ we get a special constraint on the Levi-Civita tensor:
\begin{equation}
\varepsilon^{\lambda\mu mn} \varepsilon^{\sigma\nu pq} -
\varepsilon^{\lambda\nu mn} \varepsilon^{\sigma\mu pq} =
\varepsilon^{\sigma\lambda mn} \varepsilon^{\mu\nu pq}.
\end{equation}
Furthermore, by inserting Maxwell,s field equations of motion $(\partial_\mu
F^{\mu\lambda} = 0 )$ in eq.(3.27) we find the following  equation
\begin{equation}
\L_{x^\mu}(F^{\sigma\nu}) - L_{x^\nu}(F^{\sigma\mu}) =
L_{x^\sigma}(F^{\mu\nu})
\end{equation}
which can be expressed, by using  eq.(3.20), as
\begin{equation}
\{F^{\sigma\nu},x^\mu\} - \{F^{\sigma\mu},x^\nu\} = \{F^{\mu\nu},x^\sigma\}.
\end{equation}
Also, one check that eq.(3.30) is invariant under any circular
permutation of the  indices $(\sigma ,\nu ,\mu)$ and gives the same
equation (3.28) by replacing $F^{\mu\nu}$ by its expression in terms of
$^*F^{\mu\nu}$. On the other hand, eqs.(3.27,29) give the following relation
\begin{equation}
\varepsilon^{\sigma\nu mn}\eta^{\mu\rho} - \varepsilon^{\sigma\mu
mn}\eta^{\nu\rho} = \varepsilon^{\mu\nu mn}\eta^{\sigma\rho},
\end{equation}
where $\eta^{\mu\nu}$ is the Minkowski space-time metric.
This means that eqs.(3.27,29) are equivalent. Indeed, by inserting again
the equations of motion in eq.(3.29) we get
\begin{equation}
F^{\lambda\mu}\partial_\mu\partial_\lambda F^{\sigma\nu} - \partial_\mu
F^{\lambda\nu}\partial_\lambda F^{\sigma\mu} = \partial_\mu
F^{\sigma\lambda}\partial_\lambda F^{\mu\nu}
\end{equation}
which reduces to
\begin{equation}
\partial_\mu\partial_\lambda(F^{\lambda\mu}F^{\sigma\nu} -
F^{\lambda\nu}F^{\sigma\mu} - F^{\sigma\lambda}F^{\mu\nu}) = 0
\end{equation}
and gives again eq.(3.27) up to the term $\Lambda^{\mu\lambda} =
x^\lambda C^\mu + x^\mu C^\lambda + D^{\mu\lambda}$ such that,
$\partial_\lambda D^{\mu\lambda} = 0 = \partial_\lambda D^{\mu\lambda}$ and
$C^\mu$ is a constant vector. Hence, the eq.(3.27) can be
generated, up to a quadratic term in $x^\lambda$, by inserting the
equations of motion in eq.(3.33) and so on.
\section{Symplectic form of the electromagnetic field equations of
motion}
The equations of motion of the electromagnetic field are given by
\begin{equation}
\partial_\mu F^{\mu\nu} = 0.
\end{equation}
Their dual form is the Bianchi identity
\begin{equation}
\partial_\mu \ ^*F^{\mu\nu} = 0
\end{equation}
with $^*F^{\mu\nu} = {1\over2}\varepsilon^{\mu\nu\sigma\lambda}
F_{\sigma\lambda}$ is the dual of the electromagnetic teonsor.
Furthermore, by using the Poisson bracket expression (eq.(3.4)) in
eq.(4.1) we get the geometrical form of the equations of motion:
\begin{equation}
\partial_\mu\{x^\mu,x^\nu\} = 0
\end{equation}
which can be deduced from the general formula
\begin{equation}
(2 + x^\mu\partial_\mu)\{f,x^\lambda\} = \partial_\sigma\{x^\sigma
f,x^\lambda\},
\end{equation}
where $f\in C^\infty (M_4)$, by setting $f = 1$.
The latter equation is established by using eqs.(3.21) and (2.8).
Equivalently, eq.(4.3) can be expressed as follows
\begin{equation}
[\partial_\lambda,L_{x^\lambda}] = 0 \leftrightarrow \partial_\lambda
L_{x^\lambda} = 0,
\end{equation}
where $L_{x^\lambda}$ are the symplectic diffeomorphism generators given
in eq.(3.22), because $L_{x^\lambda}$ is proportional to
$\partial_\lambda$ : $L_{x^\lambda} = F^{\sigma\lambda}\partial_\sigma$.
However, the Bianchi identity can be written as
\begin{equation}
[\partial_\lambda, ^*L_{x^\lambda}] = 0 \leftrightarrow
\partial_\lambda \ ^*L_{x^\lambda} = 0,
\end{equation}
where $^*L_{x^\lambda}\equiv \ ^*F^{\sigma\lambda}\partial_\sigma$.
Furthermore, one can verify that
\begin{equation}
L_{x^\mu}( \ ^*F_{\mu\nu}) =  \ ^*L_{x^\mu}(F_{\mu\nu}),
 \end{equation}
 $$
= \partial_\nu(\sqrt{det(F_{\mu\nu})})
$$
( by using eq.(3.8) ) and
\begin{equation}
[L_{x^\lambda}, \ ^*L_{x^\sigma}] = \partial_m (F^{m\lambda}
\ ^*F^{\mu\sigma} - F^{\mu\lambda} \ ^*F^{m\sigma})\partial_\mu
\end{equation}
which gives
\begin{equation}
[L_{x^\lambda}, \ ^*L_{x^\lambda}] = 0.
\end{equation}
The last equation can be understood from eqs.(4.5,6). Then, the equations of motion and the Bianchi identity are realized as
the electromagnetic symplectic diffeomorphisms that commute  with the
partial derivative.

Now, let us consider the Maxwell,s action
\begin{equation}
S_{A} = {-1\over4} \int_{M^{4}} {d^4x F_{\mu\nu} F^{\mu\nu}}.
\end{equation}
By replacing $F_{\mu\nu}$ by its expression (eq.(3.4)), the action
(4.10) becomes
\begin{equation}
S_{A} = {-1\over4} \int_{M^{4}} \{x_\mu,x_\nu\}\{x^\mu,x^\nu\}.
\end{equation}
It is anologous to the string action given by Schild [14] which is the
square of the usual Lagrangian representing the surface area element,
and hence does not have a purely geometrical meaning. Furthermore, by
using the following expression for the Lagrangian
\begin{eqnarray}
-2{\cal L}_A &=& - A_\lambda\partial_\mu F^{\mu\lambda} + \partial_\mu (A_\lambda
F^{\mu\lambda})\nonumber\\
&=& \{A_\lambda,x^\lambda\},
\end{eqnarray}
the action (4.11) takes the simple form
\begin{equation}
S_A = {-1\over2}\int_{M^4} \{A_\lambda,x^\lambda\}
\end{equation}
which can be rewritten (when using eq.(3.21) ) as
\begin{equation}
S_A = {-1\over2} \int_{M^4} L_{x^\lambda} (A_\lambda).
\end{equation}
It is easy to get from eqs,(4.11,14) the following identity for the
symplectic diffeomorphism generators
\begin{equation}
L_{x^\nu} (x^\mu) L_{x_\nu} (x_\mu) = 2 L_{x^\lambda} (A_\lambda).
\end{equation}

In the expression (4.13) of the action gauge invariance is manifest.
Indeed, let us consider a $U(1)$ gauge transformation for the gauge
field,
\begin{equation}
A_\lambda \rightarrow A'_\lambda = A_\lambda + \partial_\lambda
\end{equation}
then,
\begin{equation}
\{A'_\lambda,x^\lambda\} = \{A_\lambda,x^\lambda\}
\end{equation}
because we have
\begin{eqnarray}
\{\partial_\lambda\Lambda,x^\lambda\} &=&
F^{\sigma\lambda}\partial_\sigma\partial_\lambda \Lambda\nonumber\\
&=& 0
\end{eqnarray}
\section{Symplectic structure of the electromagnetic duality}
In the ref.[7] the authors have shown that the electromagnetic duality
in four dimensional space-time is a vector-vector duality: the gauge
field $A_\mu$ is interchanged with the vector field $\Lambda{_\mu}$ such
that the variation of the parent action
\begin{eqnarray}
S_{F,\Lambda} &=& \int_{M^4} d^4x ({-1\over4} F_{\mu\nu}F^{\mu\nu} +
\Lambda_\mu\partial_\nu \ ^*F^{\nu\mu})\nonumber\\
&=& \int_{M^4} d^4x ({-1\over2} L_{x^\lambda}(A_\lambda) +
\Lambda_\mu\partial_\nu \ ^*L_{x^\mu}(x^\nu))
\end{eqnarray}
with respect to $\Lambda_\mu$ gives the Bianchi identity $(\partial_\nu
\ ^*L_{x^\mu}(x^\nu))$ and the action for the electromagnetic field.
However, the variation of the same action with respect to $F_{\mu\nu}$
gives the equation
\begin{equation}
F^{\mu\nu} = -  \partial_\rho \Lambda_\sigma \varepsilon^{\rho\sigma\mu\nu}
\equiv  - \ ^*G^{\mu\nu}
\end{equation}
which can be rewritten in terms of the operators $L_{x^\mu}$ and
$\ ^*L_{x^\mu}$ as follows:
\begin{equation}
L_{x^\lambda}(A_\lambda) = - \ ^*L_{x^\lambda}(\Lambda_\lambda)
\end{equation}
and then leads to the action of the field $\Lambda_\mu$
\begin{equation}
S_{\Lambda} = - {1\over2} \int_{M^4} d^4x \ ^*L_{x^\lambda}(\Lambda).
\end{equation}
It is easy to verify that from eqs.((4.14),(5.3,4)) we get
\begin{equation}
{\cal {L}}_{A} = - {\cal{L}}_{\Lambda}
\end{equation}
which means that the two actions describe the same physics but
with different descriptions. This is the electromagnetic duality.
Furthermore, if we introduce the constant couplings $g^2$ for $A_\mu$
and $g'^2$ for $\Lambda_{\mu}$ we obtain
\begin{equation}
g'^2 = {1\over g^2}.
\end{equation}
In fact, the eq.(5.3) reflects the duality invariance between
$A_{\lambda}$ and $\Lambda_\lambda$ as follows:
We define
\begin{eqnarray}
^*(A_{\lambda}) &\equiv& \Lambda_{\lambda} ({\rm dual \ of}
A_{\lambda})\nonumber\\
^*(L_{x^\lambda}) &\equiv& \ ^*L_{x^\lambda} ({\rm dual \ of} L_{x^\lambda}),
\end{eqnarray}
and then we have
\begin{eqnarray}
^*(L_{x^\lambda}(A_{\lambda})) &=&
\ ^*L_{x^\lambda}(\Lambda_{\lambda})\nonumber\\
&=& -\ L_{x^\lambda}(A_{\lambda})
\end{eqnarray}
which means that the eq.(5.3) is  invariant with respect to  duality
transformations. On the other hand, let us consider an other local
coordinates system $(\sigma^{\mu})$ of the space -time manifold defined
by
\begin{equation}
\{\sigma^{\mu},\sigma^{\nu}\} = G^{\mu\nu} \equiv \partial^{[\mu}
\Lambda^{\nu]}
\end{equation}
to which we can associate the new symplectic diffeomorphism operators
\begin{eqnarray}
d_{\sigma^{\mu}}(\sigma^{\nu}) &\equiv & - G^{\mu\nu}\nonumber\\
^*d_{\sigma^{\mu}}(\sigma{^\nu}) &\equiv & - \ ^*G^{\mu\nu}.
\end{eqnarray}
Then, the duality condition (5.2) (or (5.3)) is rewritten as
\begin{eqnarray}
^*L_{x^\mu}(x^\nu) &=&  d_{\sigma^{\mu}}(\sigma^{\nu})\nonumber\\
L_{x^\mu}(x^\nu) &=& -\ ^*d_{\sigma{\mu}}(\sigma^{\nu})
\end{eqnarray}
where,
\begin{eqnarray}
^*L_{x^\mu}(x^\nu) &=& {1\over2} \varepsilon^{\mu\nu mn}
L_{x_m}(x_n)\nonumber\\
^*d_{\sigma^{\mu}}(\sigma^{\nu}) &=& {1\over2} \varepsilon^{\mu\nu mn}
d_{\sigma_m}(\sigma_n).
\end{eqnarray}
the constraint (5.11) (and its dual) expresses the fact that the local
system $(x^\mu)$, where the electromagnetic tensor is the electromagnetic
symplectic 2-form, is dual to the local one $(\sigma^{\mu})$, where the
tensor $G^{\mu\nu}$ defines a symplectic structure dual to the
electromagnetic one. Furthermore, the constraint (5.11) can be expressed
in terms of local coordinates by the relation
\begin{equation}
\varepsilon_{\mu\nu\lambda\rho} =  \partial_{\mu}
\sigma_{\lambda}\partial_{\nu} \sigma_{\rho},
\end{equation}
where $\sigma^{\mu}(x)$ are functions of the coordinates $(x^{\mu})$ and
$\partial_{\mu} = {\partial \over {\partial^{x^\mu}}}$. This is the
geometrical expression of the electromagnetic duality. It can be
expressed as a relationship between the operators $L_{x^{\lambda}}$ and
$d_{\sigma^{\lambda}}$ as follows:
\begin{equation}
{2\over3} \varepsilon^{\mu\nu mn} = {\varepsilon^{\mu\nu\lambda\rho} \over {\sqrt {det(F_{\mu\nu})}}}
d_{\sigma_{\rho}}(\sigma_{\lambda}) L_{x^{n}}(x^{m}),
\end{equation}
where
\begin{equation}
det(F_{\mu\nu}) \neq 0
\end{equation}
or
\begin{eqnarray}
{2\over3} \varepsilon^{\mu\nu mn} &=& {\varepsilon^{\mu\nu\lambda\rho} \over \sqrt
{det(F_{\mu\nu})}} d_{\sigma_{\rho}}(\sigma_{\lambda}) \
^*d_{\sigma^{m}}(\sigma^{n}),\nonumber\\
{2\over3}\varepsilon^{\mu\nu mn} &=& {\varepsilon^{\mu\nu\lambda\rho} \over \sqrt
{det(F_{\mu\nu})}}\ ^*L_{x_{\rho}}(x_{\lambda}) L_{x^{n}}(x^{m}).
\end{eqnarray}

To derive the equations of motion and the Bianchi identity of the two
sectors $(A_{\mu}, \Lambda_{\mu})$ we consider the constraint (5.10).
Indeed, the equations of motion for $A_{\mu}$ are given by
\begin{equation}
\partial_{\mu} L_{x^{\mu}}(x^{\nu}) = 0,
\end{equation}
and are equivalent to the equations
\begin{equation}
-\partial_{\mu} \ ^*d_{\sigma^{\mu}} (\sigma^{\nu}) = 0
\end{equation}
that express the Bianchi identity of the field $\Lambda_{\mu}$.
Inversly the equations
\begin{equation}
\partial_{\mu} \ ^*L_{x^{\mu}}(x^{\nu}) = 0
\end{equation}
express the Bianchi identity of $A_{\mu}$ and are equivalent to the
equations of motion of $\Lambda_{\mu}$:
\begin{equation}
\partial_{\mu} d_{\sigma^{\mu}}(\sigma^{\nu}).
\end{equation}
Here, by considering the duality transformations (eq.(5.3)) between two
local systems of the space-time manifold, the electromagnetic symplectic
structure transforms into the vectorial symplectic structure and
vice-versa: the gauge field in one local system transforms into a vector field
(dual to the gauge field) defined in another local system dual to the
first ( with respect to eq.(5.3)). Inversly, considering any local
system, where the vectorial symplectic structure is defined, we can
determine the local system, where the electromagnetic symplectic
structure is setting. In  other words, duality transformations
correspond to a change of coordinates associated to a special
diffeomorphisms in the sens of eq.(5.3).
\section{Conclusion}
An electromagnetic symplectic structure is established on the space-time
via the defintion of a Poisson bracket in terms of the invertible
electromagnetic tensor. furthermore, the definition of the
electromagnetic symplectic diffeomorphisms enables us to interprete
geometrically the electromagnetic duality, via the duality constraint,
by considering a dual local system of coordinates. Then, duality
transformations are understood as special diffeomorphisms on the
space-time.

It is interesting to generalize our formalism to the Yang-Mills thoery
and to its $N = 2$ supersymmetric version where duality has been proved
recently [9].

\section{ Acknowledgement}
I would like to thank Professor S.Randjbar-Daemi for his frutful discussion and I
am thankful to Professor I.Bandos for his imporatnt informations and
references. I would like to thank also Professor M.A. Virasoro for his
Hospitality at ICTP where this work was done.

\newpage
\section*{References}
\noindent \ [1] S.P. Novikov, Solitons and geometry; PISA 1994.

\noindent \ [2] L.Bars, C.Pope and E. Sezgin, Phys. Lett.B198 (1987) 455.

\noindent \ [3] F.Floratos, J. Iliopoulos and G.Tiktopoulos, Phys. Lett.
B217 (1989)285.

\noindent \ [4] M.Kachkachi and E.H. Saidi, IJMPA Vol.7 No 28 (1992)6995.

\noindent \ [5] V. Guillemin and S. Sternberg, Symplectic techniques in
physics; Cambridge University

 Press, 1988

\noindent \ [6]R. Dijkgraaf, Les Houches Lectures on fields, strings and duality; hep-th/9703136.

\noindent \ [7] S.E. Hjemeland and U. Lindstrom, Duality for non-specialists; hep-th/9705122.

\noindent \ [8] D. Olive,Phys. Lett. B72 (1997) 117

\noindent \ [9] N. Seiberg and E. Witten, hep-th/9407087.

\noindent [10] A.Sen, Phys. Lett. B329 (1994) 217.

\noindent [11] H.Kachkachi and M.Kachkachi, JMP 35 Vol.35 No.9 (1994)4477.

\noindent [12] A.T. Fomenko, Symplectic geometry; 2nd ed. Eds. Gordon \&
Breach (Luxembourg)

 1995.

\noindent [13] E. Sezgin and E. Sokatchev, ICTP preprint IC/89/72.

\noindent [14] A. Schild, Phys. Rev. Lett. B82 (1979) 247

\noindent [15] M. Monastyrsky, Topology of gauge fields and condensed
matter; Plenum Press, New

York and London, Mir publishers, Moscow, 1995.

\noindent [16] E.Witten, Conference on TQFT, Trieste, Italy, Eds. W.
Nahm, S. Randjbar-Daemi,

 E.Sezgin and E. Witten (World Scientific 1991).
\end{document}